\documentclass[12pt,reqno]{article} \usepackage{amsmath,hyperref,amssymb} \usepackage{graphicx,url}

0mm \oddsidemargin 0mm \evensidemargin 0mm \parindent 0em \parskip 1em \allowdisplaybreaks \numberwithin{equation}{section}
\newcommand{\boxedeqn}[1]{%
  \fbox{%
      \addtolength{\linewidth}{-2\fboxsep}%
      \addtolength{\linewidth}{-2\fboxrule}%
      \begin{minipage}{\linewidth}%
      \begin{align}\begin{split}#1\end{split}\end{align}%
      \end{minipage}%
    }%
} \DeclareSymbolFont{AMSa}{U}{msa}{m}{n} \DeclareSymbolFont{AMSb}{U}{msb}{m}{n} \DeclareMathSymbol{\fieldR}{\mathalpha}{AMSb}{"52}
    
\begin{document}

\begin{flushright} \small ITP--UU--11/37 \\ SPIN--11/28 \end{flushright} \bigskip \begin{center}
 {\large\bfseries On BPS bounds in D=4 N=2 gauged supergravity}\\[5mm]
Kiril Hristov$^{*,\dag}$, Chiara Toldo$^*$, Stefan Vandoren$^*$ \\[3mm]
 {\small\slshape
 * Institute for Theoretical Physics \emph{and} Spinoza Institute, \\
 Utrecht University, 3508 TD Utrecht, The Netherlands \\
\medskip
 \dag Faculty of Physics, Sofia University, Sofia 1164, Bulgaria\\
\medskip
 {\upshape\ttfamily K.P.Hristov, C.Toldo, S.J.G.Vandoren@uu.nl}\\[3mm]}
\end{center} \vspace{5mm} \hrule\bigskip \centerline{\bfseries Abstract} \medskip We determine the BPS bounds in minimal gauged supergravity in
four spacetime dimensions. We concentrate on asymptotically anti-de Sitter (AdS) spacetimes, and find that there exist two disconnected  BPS
ground states of the theory, depending on the presence of magnetic charge. Each of these ground states comes with a different superalgebra and a
different BPS bound, which we derive. As a byproduct, we also demonstrate how the supersymmetry algebra has a built-in holographic
renormalization method to define finite conserved charges.

\bigskip \hrule\bigskip

\section{Introduction and results}\label{sect:intro}

The main motivation for our analysis comes from the following paradox. Similarly to BPS states in asymptotically flat spacetimes, the authors of
\cite{Kostelecky:1995ei} provided a BPS bound in asymptotically anti-de Sitter spacetime that in the static case reduces to:
\begin{equation}\label{bound_kost} M \geq \sqrt{Q_e^2+Q_m^2}\, . \end{equation} Supersymmetric configurations would have to saturate the bound
with $M^2=Q_e^2+Q_m^2$, for a given mass $M$, electric charge $Q_e$, and magnetic charge $Q_m$ in appropriate units. However, in  $N=2$ minimal
gauged supergravity, Romans \cite{Romans:1991nq} found two supersymmetric solutions, one of which does not saturate the BPS bound, and therefore
an apparent paradox arises. The main aim of this paper is to resolve this conflict. The resolution of the paradox will lie in understanding the
BPS ground states of gauged supergravity, the associated superalgebras, and in a proper definition of the mass in asymptotically AdS$_4$
spacetimes, as we will explain in the main body of the paper.

Minimally gauged supergravity has only two bosonic fields, the metric and the graviphoton field $A_\mu$. The most general static and spherically
symmetric solution of Einstein's equation with a negative cosmological constant and an electromagnetic field is given by (in the conventions of
\cite{Romans:1991nq}), \begin{equation}\label{form_of_metric}
    {\rm d}s^2= U^2(r)\, {\rm d}t^2 - U^{-2}(r)\, {\rm d}r^2 - r^2 ({\rm d}\theta^2 + \sin^2 \theta {\rm d}\phi^2)\,,
\end{equation} with \begin{equation}\label{prefactor2}
 U^2(r) = 1- \frac{2M}{r} +\frac{Q_e^2+Q_m^2}{r^2} +g^2r^2 \ ,
\end{equation} and with nonvanishing components of the graviphoton \begin{equation}\label{ansatz_vectors}
    A_{t} = \frac{Q_e}{r} \, ,\,\,\,\,\,\,\,\,\,\,\,\,\, A_{\phi} = -Q_m \cos\theta \,\,.
\end{equation} In this class, there are two solutions that preserve some fraction of supersymmetry \cite{Romans:1991nq}. The first one is the
so-called AdS ``electric Reissner--Nordstr\"om (RN)''  solution, for which the magnetic charge $Q_m$ is set to zero and $M = Q_e$ so that the
factor $U$ has the form: \begin{equation}\label{prefactor} U^2 = \left(1- \frac{Q_e}{r}\right)^2  +g^2r^2 \, ,\,\,\,\,\,\,\,\,\,\,\,\,\, Q_m = 0
\,\,. \end{equation} This solution preserves one half of the supersymmetries (it is 1/2 BPS). Clearly, it saturates the BPS bound
\eqref{bound_kost}. Notice that the function $U(r)$ has no zeros. Therefore, there is no horizon and the point $r=0$ is a naked
singularity\footnote{In AdS spacetimes, supersymmetry does not seem to provide a cosmic censorship, contrary to most cases in asymptotically flat
spacetimes \cite{Kallosh:1992ii}. Whether cosmic censorship in AdS$_4$ can be violated is still an open problem, see e.g. \cite{Hertog:2004gz}.
This issue however has nothing to do with the paradox or contradiction mentioned above.}. Asymptotically, for $r\rightarrow \infty$, the solution
is that of pure AdS$_4$, with cosmological constant $\Lambda=-3g^2$ in standard conventions.

The second supersymmetric solution is the so-called  ``cosmic dyon'', having zero mass $M$ but nonzero fixed magnetic charge $Q_m = \pm 1/(2g)$.
Such a solution will never satisfy the BPS bound \eqref{bound_kost}. Moreover, the electric charge $Q_e$ can take an arbitrary value:
\begin{equation}\label{prefactor_magn} M=0\,,\,\,\,\,\,\,\, Q_m=\pm1/(2g)\,,\,\,\,\,\,\,\,\,   U^2  = \left(gr+\frac{1} {2gr}  \right)^2 +
\frac{Q_e^2}{r^2} \,\,. \end{equation} Again, there is a naked singularity at $r=0$. However, asymptotically, when $r\rightarrow \infty$, the
solution does not approach pure AdS$_4$, due to the presence of the magnetic charge. Instead, the solution defines another vacuum, since $M=0$,
but this vacuum is topologically distinct from AdS$_4$ in which $M=Q_m=0$. For this reason\footnote{One may argue that ground states should not
have naked singularities. Clearly, this discussion is related to cosmic censorship in AdS, which we mentioned in the previous footnote. It is
important to disentangle this discussion from the derivation of the BPS bounds. In fact, we expect that in matter coupled supergravity, there is
a magnetic ground state without naked singularities  \cite{klemm-adsBH,Dall'Agata:2010gj,Hristov:2010ri}.}, we call this vacuum magnetic anti-de Sitter,
or mAdS$_4$.

The cosmic dyon solution is 1/4 BPS, i.e. it preserves two out of eight supercharges. For both the electric RN-AdS and the cosmic dyon, the
Killing spinors were explicitly constructed in \cite{Romans:1991nq}. The fact that the BPS bound \eqref{bound_kost} is not satisfied for the
cosmic dyon leads to an apparent contradiction since states that admit a Killing spinor should saturate the BPS bound.

In this paper we show that the cosmic dyon in fact satisfies a different BPS bound that follows from a superalgebra different from the usual
AdS$_4$ superalgebra. We will determine the new BPS bound starting from the explicit calculation of the supercharges and computing the
anticommutator\footnote{An alternative approach based on the Witten-Nester energy was proposed in \cite{unpublished}, leading to similar
conclusions.}. In summary, to state the main result of this paper, for stationary configurations, the new BPS bounds are: \begin{itemize} \item
For asymptotically AdS$_4$ solutions with vanishing magnetic charge, $Q_m=0$, the BPS bound is \begin{equation}\label{bound1} M \geq |Q_e|+g
|\vec{J}|\ , \end{equation} where $\vec{J}$ is the angular momentum. \item For asymptotically magnetic AdS$_4$ solutions with $Q_m=\pm 1/(2g)$,
the BPS bound is simply \begin{equation}\label{bound2} M \geq 0 \ , \end{equation} with unconstrained electric charge $Q_e$ and angular momentum
$\vec{J}$. \end{itemize}

Other values for the magnetic charges are not considered. The quantization condition requires it to be an integer multiple of the minimal unit,
$Q_m=n /(2g); n \in {\mathbb Z}$, but it is not known if any other supersymmetric vacua can exist with $n\neq 0,1$.

The meaning of the BPS bound is not that all solutions to the equations of motion must automatically satisfy \eqref{bound1} or \eqref{bound2}.
Rather, one constructs a physical configuration space consisting of solutions that satisfy a BPS bound like  \eqref{bound1} or \eqref{bound2}.

Our procedure also provides  a new way of defining asymptotic charges in AdS$_4$ backgrounds with automatically built-in holographic
renormalization, somewhat different than the procedure developed in \cite{Kostas}. The same technique can also be applied to non-minimal gauged
supergravity (see \cite{Hristov:2009uj} for a classification of all fully BPS vacua in $N=2$ gauged supergravities). In this case, there exist
magnetically charged BPS black holes with spherical horizons \cite{klemm-adsBH,Dall'Agata:2010gj,Hristov:2010ri}. One can also extend this to
include the black brane solutions in gauged $N=2$ supergravity that were recently constructed in \cite{Barisch:2011ui}\footnote{Another possible
extension is to spaces with different asymptotics, e.g. quotients of AdS where one can have horizons of higher genus \cite{Caldarelli:1998hg}.
Here, however, we will concentrate only at spacetimes that strictly asymptote to AdS$_4$.}.

Additional motivation to study more closely the magnetic AdS$_4$ case and its superalgebra is provided by the AdS/CFT correspondence. There are
suggestions in the literature \cite{Hartnoll} that excitations of the dual theory are relevant for condensed matter physics in the presence of
external magnetic field, e.g. quantum Hall effect and Landau level splitting at strong coupling. A better understanding of the mAdS superalgebra
could then provide us with more insights about the dual field theory.

\section{Supercurrents and charges from the Noether theorem}\label{sect:Noether procedure}

In this section, we define and determine the Noether currents in minimally gauged supergravity. The currents define conserved supercharges, and
the Poisson (or Dirac) brackets between these charges produce a superalgebra. In the next section, we derive BPS bounds from the superalgebra. We
start this section by reviewing some (well-known) facts about Noether currents for local gauge symmetries.

\subsection{Generalities}

Given a Lagrangian $\mathcal{L}(\phi,\partial_\mu\phi)$, depending on fields collectively denoted by $\phi$, we have that under general field
variations \begin{equation}\label{generic_field_variation} \delta \mathcal{L}= \sum_{\phi} \mathcal{E}_{\phi} \delta \phi + \partial_{\mu}
N^{\mu} \,\,\, , \end{equation} where $\mathcal{E}_{\phi}$ vanishes upon using the equation of motion of $\phi$ and \begin{equation}
N^\mu=\frac{\delta \mathcal{L}}{\delta (\partial_\mu \phi)}\delta\phi\ . \end{equation} Under a symmetry variation, parametrized by $\epsilon$,
the Lagrangian must transform into a total derivative, such that the action is invariant for appropriate boundary conditions,
\begin{equation}\label{specific_field_variation} \delta_{\epsilon} \mathcal{L}= \partial_{\mu} K^{\mu}_{\epsilon} \,\, . \end{equation} Combining
this with  \eqref{generic_field_variation} for symmetry variations, we obtain \begin{equation} \sum_{\phi} \mathcal{E}_{\phi} \delta_{\epsilon}
\phi = \partial_{\mu} (K^{\mu}_{\epsilon}-N^{\mu}_{\epsilon}) \, . \end{equation} From the previous expression we see that the quantity
\begin{equation} J^{\mu}_{\epsilon} \equiv K^{\mu}_{\epsilon}-N^{\mu}_{\epsilon}\ , \end{equation} is the (on-shell) conserved current associated
with symmetry transformations. For the case of supersymmetry, we call $J^\mu_\epsilon$ the supercurrent. It depends on the (arbitrary) parameter
$\epsilon$ and is defined up to improvement terms of the form $\partial_{\nu} I^{\mu \nu}$ where $I$ is an antisymmetric tensor, as usual for
conserved currents. The associated conserved supercharge is then \begin{equation}\label{Q_from_J} 	\mathcal{Q} \equiv \int {\rm d}^3x
\,J^0_{\epsilon}(x)  \,. \end{equation} This supercharge should also generate the supersymmetry transformations of the fields,
\begin{equation}\label{susy_variation_from_Q} 	\delta_{\epsilon} \phi = \{\mathcal{Q},\phi \}, \end{equation} via the classical Poisson (or
Dirac in case of constraints) brackets. Since the supercurrent and correspondingly the supercharge are defined up to improvement terms and
surface terms respectively, it is not directly obvious that the Noether procedure will lead to the correct supersymmetry variations using
\eqref{susy_variation_from_Q}. In practice one always has the information of the supersymmetry variations together with the supergravity
Lagrangian, so it is possible to cross check the answers and thus derive uniquely the correct expression of the supercharge. We now illustrate
this in detail for the case of minimally gauged supergravity.

\subsection{Supercharges of minimal gauged supergravity}

First we compute the supercurrent from the Lagrangian of minimal $D=4$ $N=2$ gauged supergravity following the conventions of \cite{Ortin} (which
is written in 1.5-formalism): \begin{align} \begin{split} S = \int {\rm d}^4x& \,e \big[R(e,\omega) +6g^2 + \frac2e \epsilon^{\mu \nu \rho
\sigma} \overline{\psi}_{\mu} \gamma_5 \gamma_{\nu} (\hat{\mathcal{D}}_{\rho}+igA_{\rho}\sigma^2)\psi_{\sigma}- \mathcal{F}^2 \\
 &- \frac{1}{2e}\epsilon^{\mu\nu\rho\sigma}\overline{\psi}_{\rho} \gamma_5 \sigma^2 \psi_{\sigma}(i \overline{\psi}_{\mu} \sigma^2 \psi_{\nu} -
 \frac{1}{2e}\epsilon_{\mu\nu}{}^{\tau \lambda}  \overline{\psi}_{\tau} \gamma_5 \sigma^2 \psi_{\lambda}   ) \big]\,,
\end{split} \end{align} where $\overline{\psi}= i {\psi}^{\dagger} \gamma_0, e=\sqrt{{\rm det}g_{\mu \nu}},$  \begin{equation}\label{minimal_supercovariant_derivative}
\hat{\mathcal{D}}_{\rho} = \partial_{\rho} - \frac14 \omega_{\rho}^{ab} \gamma_{ab}-\frac{i}{2} g \gamma_{\rho}\ , \end{equation} and
\begin{align} \begin{split} \mathcal{F}_{\mu \nu} & = \partial_{\mu} A_{\nu} - \partial_{\nu} A_{\mu}+ i \overline{\psi}_{\mu} \sigma^2
\psi_{\nu} -\frac{1}{2e} \epsilon_{\mu\nu}{}^{\rho \sigma} \overline{\psi}_{\rho} \gamma_5 \sigma^2 \psi_{\sigma} = \\ &= F_{\mu \nu} + i
\overline{\psi}_{\mu} \sigma^2 \psi_{\nu} -\frac{1}{2e} \epsilon_{\mu\nu}{}^{\rho \sigma} \overline{\psi}_{\rho} \gamma_5 \sigma^2
\psi_{\sigma}\,. \end{split} \end{align} The spin connection satisfies \begin{equation} {\rm d}e^a - \omega^a{}_b \wedge e^b =0 \,\,
\end{equation} for a given vielbein $e^a = e_{\mu}^a {\rm d}x^{\mu}$. The $g^2$-term in the Lagrangian is related to the presence of a negative
cosmological constant $\Lambda= -3 g^2$.

In most of our calculations, such as in the supercurrents and supercharges, we only work to lowest order in fermions since higher order terms
vanish in the expression of the (on shell) supersymmetry algebra, where we set all fermion fields to zero. The supersymmetry variations are:
\begin{equation}\label{susy_gravitino} \delta_{\epsilon} \psi_{\mu} = \widetilde{\mathcal{D}}_\mu \epsilon= (\partial_{\mu} - \frac14
\omega_{\mu}^{ab} \gamma_{ab}-\frac{i}{2} g \gamma_{\mu} + ig A_{\mu} \sigma^2 +\frac14 F_{\lambda \tau} \gamma^{\lambda \tau} \gamma_{\mu}
\sigma^2) \epsilon \,\,\,, \end{equation} \begin{equation} \delta_{\epsilon} e_{\mu}^a = -i \overline{\epsilon} \gamma^a \psi_{\mu}\,\,\,,
\end{equation} \begin{equation}\label{susy_vector} \delta_{\epsilon} A_{\mu} = -i \overline{\epsilon} \sigma^2 \psi_{\mu}\,\,\,. \end{equation}
$U(1)$ gauge transformations act on the gauge potential and on the spinors in this way: \begin{equation} A_{\mu}'= A_{\mu} + \partial_{\mu}
\alpha\,\,, \end{equation} \begin{equation} \psi_{\mu}'= e^{-ig\alpha \sigma^2} \psi_{\mu}\,\,. \end{equation} We use the conventions in which
all the spinors are real Majorana ones\footnote{In our conventions, the two real gravitini in the gravity multiplet are packaged together in the
notation: \begin{equation} \psi_{\mu}=\left( \begin{array}{c}
 \psi_{\mu}{}^1 \\
\psi_{\mu}{}^2  \\ \end{array} \right)\, , \end{equation} where each gravitino is itself a 4-component Majorana spinor. Similar conventions are
used for the supersymmetry parameters. In other words, the $SU(2)_R$ indices are completely suppressed in our notation. }, and the gamma matrix
conventions and identities of appendix \ref{app:A}.

The quantities $N^{\mu}$ and $K^{\mu}$ for this theory are: \begin{equation} 	N^{\mu} = \frac {\partial\mathcal{L}}{\partial_{\mu}\omega}
\delta\omega+ 2 \epsilon^{\mu \nu \rho \sigma} \overline{\psi}_{\nu} \gamma_5 \gamma_{\rho} \widetilde{\mathcal{D}}_{\sigma} \epsilon  +4 i e
F^{\mu\nu}\overline{\epsilon}\sigma^2 \psi_{\nu} \,\,. \end{equation} \begin{equation} 	K^{\mu} = \frac
{\partial\mathcal{L}}{\partial_{\mu}\omega} \delta\omega - 2 \epsilon^{\mu \nu \rho \sigma} \overline{\psi}_{\nu} \gamma_5 \gamma_{\rho}
\widetilde{\mathcal{D}}_{\sigma} \epsilon  +4 i e F^{\mu\nu}\overline{\epsilon}\sigma^2 \psi_{\nu} \,\,. \end{equation} Hence the supercurrent
has the form: \begin{equation}\label{scurrent} 	J^{\mu} = - 4 \epsilon^{\mu \nu \rho \sigma} \overline{\psi}_{\nu} \gamma_5 \gamma_{\rho}
\widetilde{\mathcal{D}}_{\sigma} \epsilon  \,\,. \end{equation} This expression is gauge invariant due to the cancelation between the variation
of the gravitino, the vector field and the supersymmetry parameter. Furthermore we can show that the supercurrent is conserved
\begin{equation}
        \partial_{\mu}J^{\mu} = \partial_{\mu} (-4 \epsilon^{\mu \nu \rho \sigma}
\overline{\psi}_{\nu} \gamma_5 \gamma_{\rho} \widetilde{\mathcal{D}}_{\sigma} \epsilon) = 0 \,
\end{equation}
if we enforce the equation of motion for $\overline{\psi}_{\nu}$ and use the antisymmetry of the Levi-Civita symbol.

The Dirac brackets defined for the given theory read (we only need those containing gravitinos): \begin{equation} 	\{
\psi_{\mu}(x),{\psi}_{\sigma}(x') \}_{t=t'}= 0\,\,, \end{equation} \begin{equation} 	\{ \overline{\psi}_{\mu}(x), \overline{\psi}_{\sigma}(x')
\}_{t=t'}= 0\,\,, \end{equation} \begin{equation}\label{conjugate} 	\{ \psi_{\mu}(x), 2 \epsilon^{0 \nu \rho \sigma} \overline{\psi}_{\rho}(x')
\gamma_5 \gamma_{\sigma}
 \}_{t=t'}= \delta_{\mu}{}^{\nu} \delta^3(\vec{x}-\vec{x'})\,\,.
\end{equation} We can now check if \eqref{susy_variation_from_Q} holds with the above form of the supercurrent. It turns out that, up to overall
normalization, we indeed have the right expression without any ambiguity of improvement terms.  We only need to rescale,  since the factor of $4$
in \eqref{scurrent} does not appear in the supersymmetry variations \eqref{susy_gravitino}-\eqref{susy_vector}. The supercharge is then defined
as the volume integral\footnote{For volume and surface integrals, we use the notation that
\begin{align}
{\rm d}\Sigma_\mu=\frac{1}{6 e^2}\epsilon_{\mu\nu\rho\sigma}\,{\rm d}x^\nu \wedge {\rm d}x^\rho\wedge {\rm d}x^\sigma\ ,\qquad
{\rm d}\Sigma_{\mu\nu}=\frac{1}{2 e^2}\epsilon_{\mu\nu\rho\sigma}\,{\rm d}x^\rho\wedge {\rm d}x^\sigma\ .
\end{align}}
\begin{align}\label{susy_charge}
    \mathcal{Q} \equiv 2 \int_V {\rm d} \Sigma_{\mu} \epsilon^{\mu \nu \rho \sigma} \overline{\psi}_{\sigma} \gamma_5 \gamma_{\rho}
    \widetilde{\mathcal{D}}_{\nu} \epsilon  \,\, \stackrel{e.o.m.}{=} 2 \oint_{\partial V} {\rm d} \Sigma_{\mu \nu} \epsilon^{\mu \nu \rho
    \sigma} \overline{\psi}_{\sigma} \gamma_5 \gamma_{\rho} \epsilon\ ,
\end{align} where the second equality follows from the Gauss theorem via the equations of motion (in what follows we will always deal with
classical solutions of the theory). The Dirac bracket of two supersymmetry charges is then straightforwardly derived as the supersymmetry
variation of \eqref{susy_charge}: \begin{equation}\label{basic_susy_anticommutator_minimal} \{\mathcal{Q},\mathcal{Q} \}  = 2 \oint_{\partial V}
{\rm d}\Sigma_{\mu \nu}(\epsilon^{\mu \nu \rho \sigma} \overline{\epsilon} \gamma_5 \gamma_{\rho} \widetilde{\mathcal{D}}_{\sigma} \epsilon)\ ,
\end{equation} which is again a boundary integral.

The above formula is reminiscent of the expression for the Witten-Nester energy \cite{Witten-Nester}, which has already been implicitly assumed
to generalize for supergravity applications \cite{unpublished,Argurio,Sezgin} (see also \cite{Cheng}). Thus the correspondence between BPS bounds and positivity of
Witten-Nester energy is confirmed also in the case of minimal gauged $N=2$ supergravity by our explicit calculation of the supercharge
anticommutator.

\section{Two different BPS bounds}\label{sect:2bps_bounds}

In this section, we derive two BPS bounds based on the two BPS sectors that we consider. What is relevant for the BPS bound are the properties of
the asymptotic geometries and corresponding Killing spinors. The Killing spinors of AdS$_4$ and magnetic AdS$_4$ (``cosmic monopole'') are given
in App. \ref{app:B}, see also \cite{Romans:1991nq,unpublished}. Since only the asymptotics are important, we can set $Q_e=0$ in the cosmic dyon
solution. The AdS$_4$ solution is characterized by $M=Q_e=Q_m=0$, while mAdS$_4$ by $M=0$, $Q_e=0$, $Q_m=\pm 1/(2g)$. The corresponding Killing
spinors take a very different form: \begin{equation}\label{Killing_spinors_AdS1}
    \epsilon_{AdS} = e^{\frac{i}{2} arcsinh (g r) \gamma_1} e^{\frac{i}{2} g t \gamma_0} e^{-\frac{1}{2} \theta \gamma_{1 2}}
e^{-\frac{1}{2} \varphi \gamma_{2 3}} \epsilon_0\ , \end{equation} \begin{equation}\label{Killing_spinors_magnAdS1}
    \epsilon_{mAdS} =  \frac{1}{4} \sqrt{g r + \frac{1}{2 g r}} (1 + i \gamma_1) (1\mp i \gamma_{2 3} \sigma^2) \epsilon_0\ ,
\end{equation} where $\epsilon_0$ is a doublet of constant Majorana spinors, carrying $8$ arbitrary parameters. From here we can see that AdS$_4$
is fully supersymmetric and its Killing spinors show dependence on all the four coordinates. mAdS$_4$ on the other hand is only $1/4$ BPS: its
Killing spinors satisfiy a double projection that reduces the independent components to $1/4$ and there is no angular or time dependence. We will
come back to this remarkable fact in Section \ref{sect:Superalgebra}.

The form of the Killing spinors is important because the bracket of two supercharges is a surface integral at infinity
\eqref{basic_susy_anticommutator_minimal}. Writing out the covariant derivative in \eqref{basic_susy_anticommutator_minimal}, one obtains
\begin{equation}\label{basic_susy_anticommutator_expanded_deriv} \{ \mathcal{Q},\mathcal{Q}\}  = 2 \oint_{\partial V} {\rm d}\Sigma_{\mu \nu}
\left[ \epsilon^{\mu \nu \rho \sigma} \overline{\epsilon} \gamma_5 \gamma_{\rho}(\partial_{\sigma} - \frac14 \omega_{\sigma}^{ab}
\gamma_{ab}-\frac{i}{2} g \gamma_{\sigma} + ig A_{\sigma} \sigma^2 +\frac14 F_{\lambda \tau} \gamma^{\lambda \tau} \gamma_{\sigma} \sigma^2)
\epsilon \right]\ , \end{equation} and it depends on the asymptotic value of the Killing spinors of the solution taken into consideration.
Therefore the superalgebra will be different in the two cases and there will be two different BPS bounds.

The procedure to compute the BPS bound is the following. From \eqref{susy_charge} we have a definition of the supercharges $\mathcal{Q}_{AdS}
(\epsilon_{AdS})$ and $\mathcal{Q}_{mAdS} (\epsilon_{mAdS})$. We will then make use of the following definition for the fermionic supercharges
$Q_{AdS}, Q_{mAdS}$: \begin{equation}
    \mathcal{Q}_{AdS} \equiv Q^{T}_{AdS} \epsilon_0 = \epsilon^T_0 Q_{AdS}\ , \qquad \mathcal{Q}_{mAdS} \equiv Q^{T}_{mAdS} \epsilon_0 =
    \epsilon^T_0 Q_{mAdS}\ ,
\end{equation} i.e. any spacetime and gamma matrix dependence of the bosonic supercharges $\mathcal{Q}$ is left into the corresponding fermionic
$Q$. We are thus able to strip off the arbitrary constant $\epsilon_0$ in any explicit calculations and convert the Dirac brackets for
$\mathcal{Q}$ into an anticommutator for the spinorial supercharges $Q$ that is standardly used to define the superalgebra. Therefore now we
compute the surface integrals \eqref{basic_susy_anticommutator_minimal} for the Killing spinors of AdS$_4$ and mAdS$_4$ respectively. After
stripping off the $\epsilon_0$'s, we find the anticommutator of fermionic supercharges given explicitly in terms of the other conserved charges
in the respective vacua. The BPS bound is then derived in the standard way by requiring the supersymmetry anticommutator to be positive definite,
see e.g. \cite{gauntlett_gibbons} for details.

\subsection{Asymptotically AdS$_4$ states}

We now derive the resulting supersymmetry algebra from the asymptotic spinors of AdS$_4$. For this we use the general expression
\eqref{basic_susy_anticommutator_expanded_deriv} for the Dirac brackets of the supersymmetry charges, together with the asymptotic form of the
Killing spinors, \eqref{Killing_spinors_AdS1}. Inserting the Killing spinors $\epsilon_{AdS}$ of  \eqref{Killing_spinors_AdS1} in
\eqref{basic_susy_anticommutator_expanded_deriv}, we recover something that can be written in the following form:
\begin{equation}\label{AdS_susy_anticommutator_minimal} \{ \mathcal{Q}, \mathcal{Q} \}  = -i \, \overline{\epsilon_0} ( A + B_{a} \gamma^{a} + C
\gamma^5+ D_{i j} \gamma^{i j} + E_{i} \gamma^{0 i} + F_{a} \gamma^{a 5}){\epsilon_0}\ , \end{equation} where the charges $A,B,...$ can be
written down explicitly from the surface integral \eqref{basic_susy_anticommutator_expanded_deriv}. They will define the electric charge ($A$),
momentum ($B$), angular momentum ($D$, with $i,j=1,2,3$ spatial indices), and boost charges ($E$). The charge $C$ would correspond to a magnetic
charge, which we assumed to vanish by construction. Without the charge $F_a$, the above bracket will fit in the $Osp(2|4)$ superalgebra (see more
below). We will therefore take as definition of asymptotically AdS solutions  the ones for which $F_{a}$ vanishes. This choice of fall-off
conditions is similar to the case of $N=1$ supergravity, where the asymptotic charges are required to generate the $Osp(1|4)$ superalgebra
\cite{Teitelboim}. Extensions of the $N=2$ superalgebra where the charges $C$ and $F_a$ are non-zero have been discussed in \cite{Giuseppe}.

From the previous expression we see that conserved charges like $Q_e$, $M$ et cetera will arise as surface integrals of the five terms (or their
combinations) appearing in the supercovariant derivative. We are going to see how this works analyzing each term in the supercovariant
derivative, explicitly in terms of the ansatz of the metric \eqref{form_of_metric} and vector fields \eqref{ansatz_vectors}. This will provide us
with a new definition of the asymptotic charges in AdS$_4$ with no need to use the holographic renormalization procedure anywhere. As an explicit
example one can directly read off the definition of mass $M \equiv B_0/(8\pi)$ from the explicit form of the asymptotic Killing spinors. In the
stationary case,

\vspace{0,5cm} \boxedeqn{ \label{mass_ads} M &= \frac{1}{8\pi}\lim_{r\rightarrow\infty} \oint e\ {\rm d}\Sigma_{t r} \{ e^t_{[ 0} e^r_1
e^{\theta}_{2]} + \sin \theta e^t_{[ 0} e^r_1 e^{\varphi}_{3]} \\ &+2 g^2 r e^t_{[0} e^r_{1]} - \sqrt{g^2 r^2 +1} (\omega_{\theta}^{a b} e^t_{[ 0}
e^r_a e^{\theta}_{b]} + \omega_{\varphi}^{a b} e^t_{[ 0} e^r_a e^{\varphi}_{b]} ) \}\ .\\\\ }

We are going to take into consideration both static and rotating solutions, but we will carry out our procedure and explain the calculation in
full detail only the case of the electric RN-AdS black hole and comment more briefly on the rotating generalizations.

\begin{itemize}
  \item Electric RN-AdS\\

Here we take into consideration solutions of the form \eqref{form_of_metric} - \eqref{ansatz_vectors} with zero magnetic charge, $Q_m=0$. We
now evaluate the various terms in \eqref{basic_susy_anticommutator_expanded_deriv}.

To begin, it is easy to determine the piece concerning the field strength, namely \begin{equation}\label{basic_susy_anticommu} 2\oint {\rm
d}\Sigma_{\mu \nu} \left[\epsilon^{\mu \nu \rho \sigma} \overline{\epsilon}(t,r,\theta,\varphi) \gamma_5 \gamma_{\rho} \frac14 F_{\lambda
\tau} \gamma^{\lambda \tau} \gamma_{\sigma} \sigma^2 \epsilon(t,r,\theta,\varphi) \right]\ . \end{equation} Inserting the Killing spinors for AdS$_4$ described in \eqref{Killing_spinors_AdS1}, and exploiting the Clifford algebra relations we get
$$
2\oint {\rm d}\Sigma_{tr} \, \overline{\epsilon}_{AdS}(t, r,\theta,\varphi) e \, F^{tr} \sigma^2 \epsilon_{AdS} (t,r,\theta,\varphi)\ =
$$
$$
2\oint {\rm d}\Sigma_{tr} {\epsilon_0}^{T} e^{\frac12 \varphi \gamma_{23}} e^{\frac12 \theta \gamma_{12}}
e^{-i g t \gamma_{0}}  e^{\frac{i}{2} arcsinh(gr) \gamma_{1}} \gamma_0 e \, F^{tr} \sigma^2
e^{\frac{i}{2} arcsinh(gr) \gamma_{1}} e^{i g t \gamma_{0}}
e^{-\frac12 \theta \gamma_{12}}  e^{-\frac12 \varphi \gamma_{23}} {\epsilon_0} =
$$ \begin{equation}\label{AdS_strippedQ} = 8 i \pi \overline{\epsilon_0} Q_e \sigma^2 {\epsilon_0} \ ,
\end{equation} with the definition of the electric charge
\begin{equation}
Q_e= \frac{1}{4\pi}\oint_{S^2}F^{tr} r^2 \sin \theta\ {\rm d}\theta {\rm d}\phi \ .
\end{equation}

From here we can identify the term $A = - 8 \pi Q_e \sigma^2$ appearing in \eqref{AdS_susy_anticommutator_minimal}.

Next we consider the term containing the ``bare" gauge field $A_{\mu}$. This gives a possible contribution to the charge $F_a$ in
\eqref{AdS_susy_anticommutator_minimal}. As we mentioned above, we assumed this contribution to vanish for asymptotically AdS solutions. One
can explicitly check this for the class of electric RN-AdS solutions given in \eqref{form_of_metric}, since the only nonzero component of the
vector field is $A_t$ (see \eqref{ansatz_vectors}), hence \begin{equation}\label{A_term} \oint {\rm d}\Sigma_{tr} \left[ \epsilon^{tr \rho
\sigma} \overline{\epsilon}_{AdS}(t,r,\theta,\varphi) \gamma_5 \gamma_{\rho}  i g A_{\sigma} \sigma^2 \epsilon_{AdS}(t,r,\theta,\varphi)
\right] =0\,. \end{equation}

The term with the partial derivative $\partial_{\sigma} $ in \eqref{basic_susy_anticommutator_expanded_deriv} in the supercovariant
derivative gives nonvanishing contributions for $\sigma=\theta, \varphi$ and  it amounts to the integral:
\begin{equation}\label{AdS_strippedderivative} 2\oint {\rm d}\Sigma_{tr} \left[ \epsilon^{tr \rho \sigma}
\overline{\epsilon}_{AdS}(t,r,\theta,\varphi) \gamma_5 \gamma_{\rho}  \partial_{\sigma} \epsilon_{AdS}(t,r,\theta,\varphi) \right] = -2 i
\oint  r \overline{\epsilon_0}  \gamma_0 {\epsilon_0}  \sin \theta \, {\rm d}\theta {\rm d}\varphi \ . \end{equation} Clearly, this term will
contribute, together with other terms, to the mass.

The integral containing the spin connection is:
$$
-\frac24 \oint {\rm d}\Sigma_{tr} \epsilon^{t r \rho \sigma}   \overline{\epsilon}_{AdS} (t,r,\theta,\varphi) \gamma_5 \gamma_{\rho} \, \omega_{\sigma}^{ab} \, \gamma_{ab} \, \epsilon_{AdS}(t,r,\theta,\varphi) =
$$ \begin{equation}\label{AdS_strippedmass} = 2 i \oint \overline{\epsilon_0} \gamma^0  r
\sqrt{1+g^2r^2} \sqrt{1 +g^2r^2-\frac{2M}{r} +\frac{Q_e^2}{r^2}}\, {\epsilon_0} \sin \theta \, {\rm d}\theta {\rm d}\varphi \,,
\end{equation} where we have used \begin{equation}\label{sinh_relation} e^{ i\, arcsinh(gr) \gamma_{1}} = \sqrt{1+g^2r^2} +ig\gamma_1 r \,,
\end{equation} and the value of the spin connection: \begin{equation}\label{omega_AdS1} \omega_t^{0 1} = U \partial_r U, \quad
\omega_{\theta}^{1 2} = - U , \quad \omega_{\varphi}^{13} = - U \sin \theta, \quad \omega_{\varphi}^{23} = - \cos \theta\ . \end{equation}
Also \eqref{AdS_strippedmass} will contribute to $B^0$, and therefore to the mass.

The last contribution of the supercovariant derivative, the term proportional to $g\gamma_\sigma$, yields
\begin{equation}\label{contrib_gamma_explicit}
-2\oint {\rm d}\Sigma_{tr} \epsilon^{t r \rho \sigma} \overline{\epsilon}_{AdS}(t,r,\theta,\varphi) \frac{ig}{2} \gamma_5 \gamma_{\rho} \gamma_{\sigma} \, \epsilon_{AdS}(t,r,\theta,\varphi) =
-2 i \oint \overline{\epsilon_0} \gamma^0 r^3 g^2 {\epsilon_0}\,  \sin \theta \, {\rm d}\theta {\rm d}\varphi\ .
\end{equation} In deriving this we have used the formula $ \gamma^{tr \mu} \gamma_{\mu} = 2
\gamma^{tr}$. Again, this term contributes to the mass formula.

Collecting all the terms that contribute to the mass (the derivative term \eqref{AdS_strippedderivative}, the sum of the spin connection term
\eqref{AdS_strippedmass}, and the gamma term \eqref{contrib_gamma_explicit}) gives rise to: \begin{equation}\label{contrib_total_mass}
 2i \oint  \overline{\epsilon_0} \gamma^0 \left[ r \sqrt{1+g^2r^2} \sqrt{1 +g^2r^2-\frac{2M}{r}
+\frac{Q_e^2}{r^2}}- r^3 g^2- r \right] {\epsilon_0}\, \sin \theta \, {\rm d}\theta {\rm d}\varphi\,. \end{equation} The integral has to be
performed on a sphere with $r \rightarrow \infty$. Taking this limit one can  see that in this expression all the positive powers of $r$ are
canceled. Hence all possible divergences cancel out, and we are left with a finite contribution. In this sense, our method provides a
holographic renormalization of the mass. In the cases we can compare, our method agrees with previously known results.

Performing the integral on the remaining finite part we find: \begin{equation}\label{AdS_mass_cancellation} -8i \pi \overline{\epsilon_0} M
\gamma^{0} {\epsilon_0} = -i \overline{\epsilon_0}\gamma^0  B_0 \, {\epsilon_0} \,. \end{equation} To sum up, for the electric RN-AdS
solution, the brackets between supercharges read: \begin{align}\label{AdS_susy_anticommutator_minimal_AdSRN} \begin{split} \{ \mathcal{Q},
\mathcal{Q} \} = -8 \pi i\overline{\epsilon_0} (M \gamma^{0} - Q_e \sigma^2 ){\epsilon_0}\, \\ \Rightarrow \{\epsilon^T_0 Q, Q^T \epsilon_0
\} = 8 \pi \epsilon_0^{T} (M - Q_e \gamma^0 \sigma^2)\ \epsilon_0\ . \end{split} \end{align} Now we can strip off the constant linearly
independent doublet of spinors $\epsilon_0$ on both sides of the above formula to restore the original $SO(2)$ and spinor indices:
\begin{equation}\label{AdS_susy_anticommutator_minimal_AdSRN_indices} \{ Q^{A \alpha}, Q^{B \beta} \} = 8 \pi \left( M \delta^{AB}
\delta^{\alpha \beta} - i\,Q_e \epsilon^{AB} (\gamma^0)^{\alpha \beta}\right)\ . \end{equation} This expression coincides with the one
expected from the algebra $Osp(2|4)$ (see \eqref{osp24} in the next section) if we identify $M_{-1 0} = 8 \pi M$ and $T^{1 2} = 8 \pi Q_e$.

The BPS bound for the electric RN-AdS solution is then\footnote{See e.g. \cite{gauntlett_gibbons} for details on the general procedure of
deriving of BPS bounds from the superalgebra.}: \begin{equation}\label{BPS_bound_AdSRN} M \geq |Q_e|\ . \end{equation} The state that
saturates this bound, for which $M=|Q_e|$, preserves half of the supersymmetries, i.e. it is half-BPS. It is the ground state allowed by
\eqref{BPS_bound_AdSRN} and represents a naked singularity. All the excited states have higher mass and are either naked singularities or
genuine black holes.

It is interesting to look at the case of extremal black holes, in which inner and outer horizon coincide. This yields a relation between the
mass and charge, which can be derived from the solution given in \eqref{form_of_metric} and \eqref{prefactor2}. Explicit calculation gives
the following result \cite{Caldarelli:1998hg}:
  \begin{equation}\label{extremality_bound_RNAdS}
    M_{extr} = \frac{1}{3 \sqrt{6} g} (\sqrt{1+12 g^2 Q_e^2} +2) (\sqrt{1+12 g^2 Q_e^2} - 1)^{1/2}\ .
  \end{equation}
This lies above the BPS bound unless $Q_e = 0$, in which case we recover the fully supersymmetric AdS$_4$ space. Thus,
  \begin{equation}\label{RNAdSBPSvsExtr}
    M_{extr}> M_{BPS} \ .
  \end{equation}

  \item Kerr-AdS\\
  The Kerr-AdS black hole is an example of a stationary spacetime without
charges but with non-vanishing angular momentum. It is most standardly written in Boyer-Lindquist-type coordinates and we refer to
\cite{Caldarelli:1998hg} for more details. More details on how to calculate the angular momenta from the anticommutator of the supercharges
can be found in App. \ref{app:C}. The BPS bound is straightforward to find also in this case, leading to
  \begin{equation}\label{kerr-AdS_BPS_bound}
    M \geq g |\vec{J}|\ ,
  \end{equation}
  where the BPS state satisfies $M = g |\vec{J}|$ and in fact corresponds to a
singular limit of the Kerr-AdS black hole because the AdS boundary needs to rotate as fast as the speed of light \cite{Taylor-Robinson}. Note
that in general for the Kerr black hole we have $|\vec{J}| = a M$, where $a$ is the rotation parameter appearing in the Kerr solution in
standard notation. Thus $M = g |\vec{J}|$ implies $a = 1/g$, which is exactly the singular case. All the excited states given by $a < 1/g$
are however proper physical states, corresponding to all the regular Kerr-AdS black holes, including the extremal one. Thus the BPS bound is
always satisfied but never saturated by any physical solution of the Kerr-AdS type,
  \begin{equation}\label{KerrAdSBPSvsExtr}
    M_{extr} > M_{BPS}\ ,
  \end{equation}
as is well-known.

  \item KN-AdS\\
  The BPS bound for Kerr-Newman-AdS (KN-AdS) black holes\footnote{See again \cite{Caldarelli:1998hg}
for more detailed description of the KN-AdS black holes.} is a bit more involved due to the presence of both electric charge and angular
momentum. We will not elaborate on the details of the calculation which is straightforward. The resulting
  BPS bound is
  \begin{equation}\label{KN-AdS_BPS_bound}
    M \geq |Q_e| + g |\vec{J}| = |Q_e| + a g M\ ,
  \end{equation}
  and the ground (BPS) state is in fact quarter-supersymmetric.
  This  BPS bound is also well-known and is described in \cite{gauntlett_gibbons}.
    The BPS bound does in general not
coincide with the extremality bound, which in the case of the KN-AdS black holes is a rather complicated expression that can be found in
\cite{Caldarelli:1998hg,Taylor-Robinson}. Interestingly, the BPS bound and the extremality bound coincide at a finite non-zero value for the
mass and charge (with $ag<1$),
  \begin{equation}\label{critical_charge_KNAdS}
    |Q_{e, crit}| \equiv \sqrt{\frac{a}{g}} \frac{1}{1 - a g}\ .
  \end{equation}
  Now we have two distinct possibilities for the relation between the BPS
state and the extremal KN-AdS black hole depending on the actual value for the electric charge (there is exactly one BPS state and exactly
one extremal black hole for any value of charge $Q_e$):
  \begin{align}\label{KNAdSBPSvsExtr}
   \begin{split}
    M_{extr} > M_{BPS}\ ,& \qquad  |Q_e| \neq |Q_{e, crit}| \ ,\\
      M_{extr} = M_{BPS}\ ,& \qquad  |Q_e| = |Q_{e, crit}|\ .
  \end{split}
  \end{align}
So for small or large enough electric charge the BPS solution will be a naked singularity and the extremal black hole will satisfy but not
saturate the BPS bound, while for the critical value of the charge the extremal black hole is supersymmetric and all non-extremal solutions
with regular horizon will satisfy the BPS bound.

\end{itemize}

\subsection{Magnetic AdS$_4$} Unlike the standard AdS$_4$ case above, the Killing spinors of magnetic AdS$_4$ already break $3/4$ of the
supersymmetry, c.f. \eqref{Killing_spinors-magnAdS}. The projection that they obey is, \begin{equation}\label{projection_magneticAdS}
    \epsilon_{mAdS} = P \epsilon_{mAdS}\ , \qquad \qquad P \equiv \frac{1}{4}(1+ i \gamma_1) (1\mp i \gamma_{2 3} \sigma^2)\ ,
\end{equation} for either the upper or lower sign, depending on the sign of the magnetic charge. Furthermore, one has the following properties of
the projection operators, \begin{align}\label{proj} \begin{split}
    P^{\dag} P&= P^{\dag} i \gamma_1 P = \pm P^{\dag}
i \gamma_{2 3} \sigma^2  P = \pm P^{\dag} (-i \gamma_0 \gamma_5 \sigma^2)  P= P\ , \end{split} \end{align} and all remaining quantities of the
form $P^{\dag} \Gamma P$ vanish, where $\Gamma$ stands for any of the other twelve basis matrices generated by the Clifford algebra.

These identities allow us to derive, from \eqref{basic_susy_anticommutator_expanded_deriv}, the bracket
\begin{equation}\label{anticommut_magnetic_AdS}
  \{ \mathcal{Q} , \mathcal{Q} \} =  \overline{P\epsilon}_{0} \gamma_0 (-i \,8 \pi )  M P \epsilon_0 \qquad
  \Rightarrow \qquad\{\epsilon^T_0 P Q, (P Q)^T \epsilon_0 \} = \epsilon^T_0 (8 \pi M) P \epsilon_0\ .
\end{equation} provided that the mass is given by

\vspace{0,5cm}

\boxedeqn{ \label{mass_mads} M &= \frac{1}{8\pi} \lim_{r\rightarrow\infty}  \oint e\ {\rm d}\Sigma_{t r} \left( g r + \frac{1}{2 g r} \right) \bigg(
2 g (A_{\theta} e^t_{[ 0} e^r_2 e^{\theta}_{3]} + A_{\varphi} e^t_{[ 0} e^r_2 e^{\varphi}_{3]}) \\ &+ \frac{\sin \theta}{g} e^t_{[ 0} e^r_1 e^{\theta}_{2}
e^{\varphi}_{3]} + 2 g e^t_{[0} e^r_{1]} - (\omega_{\theta}^{a b} e^t_{[ 0} e^r_a e^{\theta}_{b]} + \omega_{\varphi}^{a b} e^t_{[ 0} e^r_a
e^{\varphi}_{b]} ) \bigg)\ .\\\\ }

This expression simplifies further if we choose to put the vielbein matrix in an upper triangular form, such that we have nonvanishing
$e_t^{0,1,2,3}, e_r^{1,2,3}, e_\theta^{2,3}, e_\varphi^3,$ and the inverse vielbein has only components $e_0^{t,r,\theta,\varphi},
e_1^{r,\theta,\varphi}, e_2^{\theta,\varphi}, e_3^{\varphi}$. The mass is then \begin{align} M &= \frac{1}{8\pi} \lim_{r\rightarrow\infty}  \oint e\ {\rm d} \Sigma_{t r}\left( g r + \frac{1}{2 g r} \right)\left( \frac{\sin \theta}{g} e^t_{0} e^r_1 e^{\theta}_{2} e^{\varphi}_{3} +2 g e^t_{0} e^r_{1} -
(\omega_{\theta}^{1 2} e^t_{0} e^r_1 e^{\theta}_{2} + \omega_{\varphi}^{1 3} e^t_{0} e^r_1 e^{\varphi}_{2} ) \right)\ . \end{align} Notice that
this mass formula is different from the one for asymptotically AdS$_4$ spacetimes.

Stripping off the parameters $\epsilon_0$ in \eqref{anticommut_magnetic_AdS}, leaves us with a matrix equation in spinor space. Due to the
projection operators, one is effectively reducing the number of supercharges to two instead of eight. These two supercharges are scalars, since
the Killing spinors are invariant under rotation as they don't depend on the angular coordinates (see also in the next section). Denoting them by
$Q^1$ and $Q^2$,  the anticommutator then becomes \begin{equation}\label{QQ-mads} \{ Q^I , Q^J \} = 8 \pi M \delta^{IJ} \,\,,\qquad I,J=1,2\ .
\end{equation} Hence the BPS bound is just \begin{equation}\label{BPS_bound_magnetic_AdS}
  M \geq 0\, .
\end{equation} Saturating the bound leads to a quarter-BPS solution. None of the other conserved charges, i.e. the electric charge and (angular)
momentum, influences the BPS bound due to the projection relation \eqref{proj}. Thus $Q_e$ and $\vec{J}$ can be completely arbitrary.

In particular, for the case of of the Reissner-Nordstr\"om solution \eqref{form_of_metric} with fixed magnetic charge $Q_m= \pm 1/(2g)$ and $Q_e$
arbitrary, the mass integral \eqref{mass_mads} yields $$ \frac{1}{4\pi} \lim_{r\rightarrow\infty}  \oint  \left( gr+ \frac{1}{2gr} \right) \left(
r \sqrt{1 + g^2 r^2 - \frac{2M}{r}+ \frac{Q_e^2+1/(4g^2)}{r^2} }- g \,r^2- \frac{1}{2g} \right) \sin \theta \, {\rm d}\theta {\rm d}\varphi = $$
\begin{equation} = \frac{1}{4\pi} \oint \left(-g^2r^3 - \frac{r}{2} + M - \frac{r}{2} + g^2r^3 + r \right) \sin \theta \, {\rm d}\theta {\rm
d}\varphi = M \,. \end{equation} This is exactly the mass parameter $M$ appearing in \eqref{prefactor2}. The supersymmetric solutions found by
Romans (the so-called cosmic monopole/dyons) have vanishing mass parameter hence indeed saturate the BPS bound \eqref{BPS_bound_magnetic_AdS}.

Of course in the context of a rotating black hole vanishing mass results in vanishing angular momentum due to the proportionality between the
two, i.e. an asymptotically mAdS Kerr-Newman with non-zero angular momentum spacetime can never saturate the BPS bound
\eqref{BPS_bound_magnetic_AdS}. Nevertheless, excitations over the magnetic AdS$_4$ include all Reissner-Nordstr\"{o}m and Kerr-Newman AdS black
holes that have fixed magnetic charge $2 g Q_m = \pm 1$ and arbitrary (positive) mass, angular momentum and electric charge. All these solutions
satisfy the magnetic AdS$_4$ BPS bound.

\section{Superalgebras}\label{sect:Superalgebra}

\subsection{AdS$_4$}\label{sect:AdS_algebra}

The procedure we used to find the BPS bound determines also the superalgebras of AdS$_4$ and mAdS$_4$, which are found to be different. In fact,
given the Killing spinors and Killing vectors, there is a general algorithm to determine the superalgebra, see \cite{Figueroa} and chapter 13 in
\cite{Ortin_book}.

For what concerns the pure AdS$_4$, in $N=2$ gauged supergravity the superalgebra is $Osp(2|4)$, which contains as bosonic subgroup $SO(2,3)
\times SO(2)$: the first group is the isometry group of AdS$_4$ and the second one corresponds to the gauged R-symmetry group that acts by
rotating the two gravitinos. The algebra contains the  generators of the $SO(2,3)$ group $M_{MN}$ ($M,N = -1, 0,1,2,3$),  and
$T^{AB}=-T^{BA}=T\epsilon^{AB}, A,B=1,2$, the generator of  $SO(2)$. Furthermore we have supercharges $Q^{A \alpha}$ with $A=1,2$ that are
Majorana spinors. The non-vanishing (anti-)commutators of the $Osp(2|4)$ superalgebra are: \begin{align}\label{osp24} \begin{split} [Q^{A
\alpha},T] &= \epsilon^{AB} Q^{B \alpha} \\ [M_{MN},M_{PQ} ] &= - \eta_{MP} M_{NQ} - \eta_{NQ} M_{MP} + \eta_{MQ} M_{NP} +\eta_{NP} M_{MQ} \\
[Q^{A \alpha},M_{MN} ] &= \frac12 { (\hat{\gamma}_{MN})}^{\alpha}{}_{\beta} Q^{A \beta} \\ \{ Q^{A \alpha},Q^{B \beta} \} &=  \delta^{AB}
(\hat{\gamma}^{MN}C^{-1})^{\alpha \beta} M_{MN} - (C^{-1})^{\alpha\beta} T \epsilon^{AB}\ , \end{split} \end{align} where $\eta_{MN}=  {\rm
diag}(1,1,-1,-1,-1)$, the gamma matrices are $\hat{\gamma}_M\equiv \{\gamma_5,  i \gamma_{\mu} \gamma_5\}$, and ${\hat
\gamma}_{MN}=\frac{1}{2}[\hat\gamma_M,\hat\gamma_N]$. Further details can be found in \cite{Ortin_book}. $T$ does not have the role of a central
charge, as it doesn't commute with the supercharges. Nevertheless it is associated to the electric charge\footnote{If we perform a
Wigner--In\"{o}n\"{u} contraction of the algebra, $T^{AB}$ gives rise to a central charge in the Poincar\'e superalgebra. See \cite{Ortin_book}
for further details.}. The isometry group of AdS$_4$ is $SO(2,3)$, isomorphic to the conformal group in three dimensions, whose generators are 3
translations, 3 rotations, 3 special conformal transformations (conformal boosts) and the dilatation.

\subsection{mAdS$_4$}\label{sect:mAdS_algebra} In the case of mAdS$_4$, the symmetry group is reduced due to the presence of the magnetic charge.
Spatial translations and boosts are broken, because of the presence of a magnetic monopole. There are 4 Killing vectors related to the invariance
under time translations and rotations. The isometry group of this spacetime is then $ \mathbb{R} \times SO(3)$. Furthermore, we have also gauge
invariance. The projector \eqref{projection_magneticAdS} reduces the independent components of the Killing spinors to 1/4, consequently the
number of fermionic symmetries of the theory is reduced, as we explained in section 3.2. We have denoted the remaining two real supercharges with
$Q^I$ ($I=1,2$). To sum up, the symmetry generators of mAdS$_4$ are: \begin{itemize}
\item the angular momentum $ J_i $, $ i=1,2,3$, \item the Hamiltonian $H$, \item the gauge transformation generator $T$, \item the two
supercharges $ Q^I$ where $I=1,2$. \end{itemize} From \eqref{QQ-mads} the anticommutator between two supercharges is
\begin{equation}\label{magnetic_AdS_superalgebra} \{ Q^I , Q^J \} = H \delta^{IJ} \ . \end{equation} Since  mAdS$_4$ is static and spherically
symmetric, we have the commutation relations \begin{equation} [ H , J_i ] = 0\ ,\qquad [ J_i , J_j ] = \epsilon_{ijk} \,J_k \ . \end{equation}
The following commutators are then determined by imposing the Jacobi identities: \begin{equation}\label{scalar_charges} [ Q^I , J_i ] = [ Q^I , H
] = 0 \ . \end{equation} Next, we add the gauge generator $T$ to the algebra. Because of gauge invariance, we have the commutators
\begin{equation} [ T , J_i ] = [T , H ] = 0 \ . \end{equation} From the Jacobi identities one now derives that \begin{equation} [Q^I , T ] =
\epsilon^{IJ} Q^{J} \ , \end{equation} with a fixed normalization of $T$. This commutator also follows from the observation that gauge
transformations act on the supersymmetry parameters in gauged supergravity, together with the fact that $T$ commutes with the projection operator
$P$ defined in the previous section.

The first commutator in \eqref{scalar_charges} implies that the supercharges $Q^I$ are singlet under rotations. This is a consequence of the fact
that the mAdS$_4$ Killing spinors have no angular dependence \cite{Romans:1991nq}. Group theoretically, this follows from the fact that the group
of rotations entangles with the $SU(2)_R$ symmetry, as explained in \cite{bernard_maaike}.

\section{Outlook}\label{sect:conclusion}

The procedure outlined in this paper to compute the BPS bound is completely general. It can be used also in other settings, for example for
gauged supergravity with matter couplings.  This would give new insights about the BPS ground states of the gauged supergravities. For instance,
it would be interesting to investigate the BPS bounds in the presence of scalar fields that belong to vector- or hypermultiplets.

One disadvantage of our approach is that our renormalized mass formulas \eqref{mass_ads} and \eqref{mass_mads} are written in specific
coordinates and therefore not manifestly diffeomorphism invariant. It would be an improvement if a coordinate independent formulation can be
given, and comparison can be made to other proposals in the literature, such as \cite{AD,Teitelboim,Kostas,Compere}.

Furthermore, it would be worth understanding the implications related to the AdS/CFT correspondence, especially in the presence of a magnetic
charge. For instance, the fact that the supercharges carry no spin might have important consequences for the dual field theory. We leave this for
further study.

\section*{Acknowledgements}

We thank T. Ort\'{\i}n for passing us a set of unpublished notes written by him and two collaborators (P. Meessen and J.M. Izquierdo) on the topic
covered in this paper. They contain equations that coincide with some of those found by us. We would further like to thank B. de Wit, G.
Dibitetto, I. Lodato, S. Katmadas, D. Klemm and E. Sezgin for helpful discussions and correspondence. We acknowledge support by the Netherlands
Organization for Scientific Research (NWO) under the VICI grant 680-47-603.

\appendix

\section{Gamma matrix conventions}\label{app:A} The Dirac gamma-matrices in four dimensions satisfy \begin{equation}
 \{\gamma_a,\gamma_b\} = 2 \eta_{ab}\ ,
\end{equation} and we define \begin{equation}
   [\gamma_a,\gamma_b] \equiv 2 \gamma_{ab}\ ,\qquad
\gamma_5 \equiv - i \gamma_0 \gamma_1 \gamma_2 \gamma_3 = i \gamma^0\gamma^1 \gamma^2 \gamma^3\ . \end{equation} In addition, they can be chosen
such that \begin{align}
  \gamma_0^\dagger = \gamma_0, \quad \gamma_0 \gamma_i^\dagger
  \gamma_0 = \gamma_i,\quad \gamma_5^\dagger = \gamma_5,\quad   \gamma_a^* = -\gamma_a\ .
\end{align} An explicit realization of such gamma matrices is the Majorana basis, given by \begin{align}
  \gamma^0 &= \begin{pmatrix}0 & \sigma^2\\ \sigma^2 &
    0 \end{pmatrix},&
  \gamma^1 &= \begin{pmatrix}i \sigma^3 & 0\\ 0&i\sigma^3 \end{pmatrix},&
  \gamma^2 &= \begin{pmatrix}0 & -\sigma^2\\ \sigma^2 &
    0 \end{pmatrix},\nonumber\\
  \gamma^3 &= \begin{pmatrix}-i \sigma^1 & 0\\
    0&-i\sigma^1 \end{pmatrix},&
  \gamma_5 &= \begin{pmatrix}\sigma^2 & 0\\ 0&-\sigma^2 \end{pmatrix}\ ,
\end{align} where the $\sigma^i$ ($ i=1,2,3$) are the Pauli matrices, \begin{equation} \sigma^1 = \left( \begin{array}{cc} 0 & 1  \\ 1 & 0
\end{array} \right), \quad \sigma^2 = \left( \begin{array}{cc} 0 & -i  \\ i & 0  \end{array} \right), \quad \sigma^3 = \left( \begin{array}{cc} 1
& 0  \\ 0 & -1  \end{array} \right)\ . \end{equation} For the charge conjugation matrix, we choose \begin{equation} C=i \gamma^0\ ,
\end{equation} hence Majorana spinors have real components.

We also make use of the following identities, with curved indices: \begin{equation}
    \epsilon^{\mu \nu \rho \sigma} \gamma_5 \gamma_{\rho} = i e \gamma^{\mu \nu \sigma}  \,,
\end{equation} \begin{equation}
    \gamma_{\mu} \gamma_{\rho \sigma} = -\gamma_{\rho} g_{\mu \sigma}+ \gamma_{\sigma}
g_{\mu \rho}+ \frac{i}{e} \epsilon_{\mu \nu \rho \sigma} \gamma_5 \gamma^{\nu} \,, \end{equation} \begin{equation} \gamma_{\mu} \gamma_{\nu
\rho}\gamma_{\sigma} - \gamma_{\sigma} \gamma_{\nu \rho}\gamma_{\mu}= 2 g_{\mu \nu} g_{ \rho \sigma}- 2g_{\mu \rho} g_{\nu \sigma} +2 \frac{i}{e}
\epsilon_{\mu \nu \rho \sigma} \gamma_5 \,. \end{equation} Antisymmetrizations are taken with weight one half, and the totally antisymmetric Levi
Civita symbol is defined by \begin{equation} \epsilon^{0123} = 1 = - \epsilon_{0123} \,. \end{equation} With curved indices, \begin{equation}
\epsilon^{\mu\nu\rho\sigma}\equiv e\,e^\mu_ae^\nu_be^\rho_ce^\sigma_d\,\epsilon^{abcd}\ , \end{equation}
 is a tensor-density.

Another important property that ensures the super-Jacobi identities of $Osp(2|4)$ hold is \begin{equation}
    (\hat{\gamma}^{MN} C^{-1})^{\alpha \beta} (\hat{\gamma}_{MN} C^{-1})^{\gamma \delta} = (C^{-1})^{\alpha \gamma} (C^{-1})^{\beta \delta} +
    (C^{-1})^{\alpha \delta} (C^{-1})^{\beta \gamma}\ ,
\end{equation} where $\hat{\gamma}_{MN}$ are defined in section \ref{sect:AdS_algebra}.

\section{Asymptotic Killing spinors}\label{app:B}

\subsection{AdS$_4$}\label{app:AdS} Here we give details about the Killing spinors for AdS$_4$. We consider the metric in spherical coordinates
\begin{equation}\label{metric-AdS} {\rm d} s^2 = (1+ g^2 r^2)\, {\rm d}t^2 - (1+g^2 r^2)^{-1}\, {\rm d}r^2 - r^2\, ({\rm d} \theta^2 + \sin^2
\theta {\rm d} \varphi^2)\ , \end{equation} and corresponding vielbein \begin{equation}\label{vierbein_AdS} e_{\mu}^a = {\rm
diag}\Big(\sqrt{1+g^2 r^2}, \sqrt{1+g^2 r^2}^{-1}, r, r \sin \theta\Big)\ . \end{equation} The non-vanishing components of the spin connection
turn out to be: \begin{equation}\label{omega-AdS} \omega_t^{0 1} = g^2 r, \quad \omega_{\theta}^{1 2} = -\sqrt{1+g^2 r^2}, \quad
\omega_{\varphi}^{13} = - \sqrt{1+g^2 r^2} \sin \theta, \quad \omega_{\varphi}^{23} = - \cos \theta\ . \end{equation} For the AdS$_4$ solution,
the field strength vanishes, \begin{equation}\label{field_strengths-AdS} F_{\mu \nu} = 0\ . \end{equation} To find the Killing spinors
corresponding to this spacetime we need to solve $\widetilde{\mathcal{D}}_{\mu} \epsilon = 0$. This equation has already been solved in
\cite{unpublished} and we have explicitly checked that the resulting Killing spinors are given by \begin{equation}\label{Killing_spinors-AdS}
    \epsilon_{AdS} = e^{\frac{i}{2} arcsinh (g r) \gamma_1} e^{\frac{i}{2} g t \gamma_0} e^{-\frac{1}{2} \theta \gamma_{1 2}} e^{-\frac{1}{2}
    \varphi \gamma_{2 3}} \epsilon_0\ ,
\end{equation} where $\epsilon_0$ is a doublet of arbitrary constant Majorana spinors, representing the eight preserved supersymmetries of the
configuration.

It is important to note that the asymptotic solution of the Killing spinor equations as $r\rightarrow\infty$ (given the same asymptotic metric)
cannot change unless $A_{\varphi} \neq 0$. This is easy to see from the form of the supercovariant derivative
\eqref{minimal_supercovariant_derivative} since any other term would necessarily vanish in the asymptotic limit. More precisely, any gauge field
carrying an electric charge that appears in the derivative vanishes asymptotically, the only constant contribution can come when a magnetic
charge is present. In other words, any spacetime with vanishing magnetic charge and asymptotic metric \eqref{metric-AdS} has asymptotic Killing
spinors given by \eqref{Killing_spinors-AdS}.

\subsection{Magnetic AdS$_4$}\label{app:magnetic AdS} Now we will show that the asymptotic Killing spinors take a very different form when
magnetic charge is present. In this case the metric is \begin{equation}\label{metric-magnAdS} {\rm d} s^2 = (1+ g^2 r^2+\frac{Q_m^2}{r^2})\, {\rm
d}t^2 - (1+g^2 r^2+\frac{Q_m^2}{r^2})^{-1}\, {\rm d}r^2 - r^2\, ({\rm d} \theta^2 + \sin^2 \theta {\rm d} \varphi^2)\ , \end{equation} with
corresponding vielbein: \begin{equation}\label{vierbein_magnAdS} e_{\mu}^a = {\rm diag}\Big(\sqrt{1+g^2 r^2+\frac{Q_m^2}{r^2}}, \sqrt{1+g^2
r^2+\frac{Q_m^2}{r^2}}^{-1}, r, r \sin \theta\Big)\ . \end{equation} The non-vanishing components of the spin connection turn out to be:
\begin{eqnarray}\label{omega-magnAdS} &&\omega_t^{0 1} = g^2 r-\frac{Q_m^2}{r^3}\ , \qquad \omega_{\theta}^{1 2} = -\sqrt{1+g^2
r^2+\frac{Q_m^2}{r^2}}\ ,\nonumber\\ &&\omega_{\varphi}^{1 3} = - \sqrt{1+g^2 r^2+\frac{Q_m^2}{r^2}} \sin \theta\ , \qquad \omega_{\varphi}^{2 3}
= - \cos \theta\ . \end{eqnarray} As opposed to the previous section, now we have a non-vanishing gauge field component $A_{\varphi} = - Q_m \cos
\theta$, resulting in $F_{\theta \varphi} = Q_m \sin \theta$. If we require $\widetilde{\mathcal{D}}_{\mu} \epsilon = 0$ and insist that $Q_m
\neq 0$, we get a solution described by Romans in \cite{Romans:1991nq} as a ``cosmic monopole'' (which we call magnetic AdS$_4$). The magnetic
charge satisfies $2 g Q_m = \pm 1$, such that the metric function is an exact square $(g r + \frac{1}{2 g r})^2$. The Killing spinors
corresponding to solutions with $Q_m=\pm 1/(2g)$ in our conventions are given by \begin{equation}\label{Killing_spinors-magnAdS}
    \epsilon_{mAdS} =  \frac{1}{4} \sqrt{g r + \frac{1}{2 g r}} (1+ i \gamma_1) (1\mp i \gamma_{2 3} \sigma^2) \, \epsilon_0\ ,
\end{equation} preserving two of the original eight supersymmetries. Note that in the limit $r\rightarrow\infty$ the Killing spinor projections
continue to hold. Furthermore, the functional dependence is manifestly different in the expressions \eqref{Killing_spinors-AdS} and
\eqref{Killing_spinors-magnAdS} for the Killing spinors of ordinary AdS$_4$ and its magnetic version. This leads to the conclusion that these two
vacua and their corresponding excited states belong to two separate classes, i.e. they lead to two independent superalgebras and BPS bounds. Note
that one can also add an arbitrary electric charge $Q_e$ to the above solution, preserving the same amount of supersymmetry (the ``cosmic dyon''
of \cite{Romans:1991nq}). The corresponding Killing spinors \cite{Romans:1991nq} have the asymptotic form of \eqref{Killing_spinors-magnAdS},
i.e. the cosmic dyons are asymptotically magnetic AdS$_4$.

\section{Rotations in AdS$_4$}\label{app:C} Here we focus on stationary spacetimes with rotations. From thesupersymmetry Dirac brackets in
asymptotic AdS$_4$ spaces, \begin{equation} \{ \mathcal{Q}, \mathcal{Q} \} =- 8 \pi i \overline{\epsilon}_{0}\left( ...+ g J_{i j} \gamma^{i j}+
... \right)  \epsilon_{0}\ , \end{equation} we can derive a definition of the conserved angular momenta. The explicit expressions are somewhat
lengthy and assume a much simpler form once we choose the vielbein matrix $e_{\mu}^a$ in an upper triangular form, such that its inverse
$e_a^{\mu}$ is also upper triangular. More explicitly, in spherical coordinates we choose nonvanishing $e_t^{0,1,2,3}, e_r^{1,2,3},
e_\theta^{2,3}, e_\varphi^3,$ such that the inverse vielbein has only non-vanishing components $e_0^{t,r,\theta,\varphi}, e_1^{r,\theta,\varphi},
e_2^{\theta,\varphi}, e_3^{\varphi}$. The resulting expressions for the angular momenta in this case become: \begin{align}\label{ang_mom}
\nonumber    J_{12} = \frac{1}{8 \pi} \lim_{r\rightarrow\infty} \int\limits_{0}^{2 \pi} {\rm d} \varphi \int\limits_{0}^{\pi} {\rm d} \theta &
\left( (e_0^t e_1^r e_{\varphi}^3 \omega_{\theta}^{01} + e_0^t e_1^r e_{\theta}^2 e_{\varphi}^3 e_2^{\varphi} \omega_{\varphi}^{01}) r \cos
\varphi + (e_0^t e_1^r e_{\theta}^2 \omega_{\varphi}^{01}) r \cos \theta \sin \varphi \right)\ , \\ \nonumber    J_{13} = \frac{1}{8 \pi}
\lim_{r\rightarrow\infty} \int\limits_{0}^{2 \pi} {\rm d} \varphi \int\limits_{0}^{\pi} {\rm d} \theta & \left( (e_0^t e_1^r e_{\varphi}^3
\omega_{\theta}^{01} + e_0^t e_1^r e_{\theta}^2 e_{\varphi}^3 e_2^{\varphi} \omega_{\varphi}^{01}) r \sin \varphi + (e_0^t e_1^r e_{\theta}^2
\omega_{\varphi}^{01}) r \cos \theta \cos \varphi \right)\ , \\
    J_{23} &= \frac{1}{8 \pi} \lim_{r\rightarrow\infty} \int\limits_{0}^{2 \pi} {\rm d} \varphi \int\limits_{0}^{\pi} {\rm d} \theta \left( e_0^t
    e_1^r e_{\theta}^2 \omega_{\varphi}^{01} r \sin \theta \right) \ .
\end{align} It is easy to see that in case of axisymmetric solutions around $\varphi$, such as the Kerr and Kerr-Newman metrics in AdS, the
angular momenta $J_{12}$ and $J_{13}$ automatically vanish due to $\int_0^{2 \pi} {\rm d} \varphi \sin \varphi = \int_0^{2 \pi} {\rm d} \varphi
\cos \varphi = 0$.

One can then use the formula for $J_{23}$ to derive the value of the angular momentum for the Kerr black hole. This is still somewhat non-trivial
because one needs to change the coordinates from Boyer-Lindquist-type to spherical. The leading terms at large $r$ were found in appendix B of
\cite{Teitelboim} and are enough for the calculation of the angular momentum since subleading terms vanish when the limit is taken in
\eqref{ang_mom}. The calculation of the relevant component of the spin connection leads to
 \begin{equation}
    \omega_{\varphi}^{01} = - \frac{3 a m \sin^2 \theta (1-g^2 a^2 \sin^2 \theta)^{-5/2}}{r^2} + \mathcal{O} (r^{-3})
 \end{equation}
 and gives the exact same result as in (B.8) of \cite{Teitelboim},
 \begin{equation}
   J_{23} = \frac{a m}{(1-g^2 a^2)^2}\ .
 \end{equation}
This expression has also been derived from different considerations in \cite{Cognola}, thus confirming the consistency of our results.

One can also verify the result for the asymptotic mass of the Kerr and Kerr-Newman spacetimes using \eqref{mass_ads} and the metric in appendix B
of \cite{Teitelboim}. After a somewhat lengthy but straightforward calculation one finds \begin{align}
    M =  \frac{m}{(1-g^2 a^2)^2}\ ,
\end{align} as expected from previous studies (see, e.g., \cite{Caldarelli:1998hg,Cognola}).

\end{document}